\documentclass[%
aps, %这个是指APS风格，另一个可选项应该是AIP
prl,%pra, prb, rmp, % 这里是对不同期刊的选择
superscriptaddress,
reprint,
showpacs,
amsmath, amssymb, %都是用于显示数学公式环境
]{revtex4-2}% 这个格式就是APS对应的latex文档格式
\usepackage{graphicx}% Include figure files
\usepackage{dcolumn}% Align table columns on decimal point
\usepackage{bm}% bold math
\usepackage{natbib}%to cite
\usepackage{hyperref}% add hypertext capabilities
\usepackage{float}
\usepackage{mathrsfs}%花体的插入方法
\hypersetup{colorlinks=true, citecolor=blue, urlcolor=blue, linkcolor=blue}
\begin{document}
\title{Hydrating and dehydrating dynamics process as an ion entering a carbon nanotube}
\author{Zhenyu Wei}

\author{Yunfei Chen}
\email{yunfeichen@seu.edu.cn}
\affiliation{School of Mechanical Engineering, Southeast University, Nanjing 211189, China}

\begin{abstract}% 摘要
    Ion transport within confined environments, like nanopores and nanotubes, is pivotal for advancing a range of applications, including biosensors, seawater desalination, and energy storage devices. Existing research does not sufficiently delve into the alterations in ion hydration structure, a key element that significantly influences ion transport properties in such confined environments. Here, we report a theoretical model comprehensively considering the interplay between ions and nanotubes. By incorporating the effect of orientational competition, our model effectively predict both the change in hydration structure and the free energy profile during an ion's transition from bulk water into a nanotube. These predictions match closely to the results from molecular dynamics simulations.
\end{abstract}
\maketitle

% Introduction

The transport properties of ions in confined environments, like those within a carbon nanotube, are distinctly different from those in bulk conditions\cite{Kavokine2021}. This unique characteristic makes carbon nanotubes promising candidates for applications in supercapacitors\cite{Chmiola2006a, Zhang2009a,Yang2019, Wang2020} and electric circuits\cite{T.Mouterde2019,Lucas2020,Xue2021}. The primary factor influencing ion transport properties is the hydration layer, denoting the local water structure that forms around an ion. For example, the Walden's rule fails\cite{Zwanzig1963,Banerjee2019} to adequately describe ion behavior in water, largely due to the neglect of friction arising from the existence of hydration layer. Furthermore, the water structure within a confined environment diverges significantly from that in bulk conditions, tending to form an ice-like arrangement within the nanotube\cite{Koga2001, He2013a}. This is largely attributed to the formation of a unique hydrogen bond network\cite{DallaBernardina2016} in the carbon nanotube. Consequently, the hydration layers surrounding an ion is also greatly affected, leading to dramatically different transport properties. Numerous studies\cite{Xue2021, Shao2009a} have shown that different types of ions undergo varying dehydration processes when entering a carbon nanotube with a radius of less than $1\text{nm}$ that gives rise to significant free energy barriers\cite{Zwolak2009, Sahu2017c}. Given these complexities, a model capable of describing the intricate phenomena of hydrating and dehydrating dynamics is essential for understanding and harnessing the unique behavior of ion transport within confined environments.

To date, numerous studies have employed various methodologies to investigate the changes in the hydration structure around an ion inside a nanotube and its subsequent effect on ion transport properties. Ventra et al.\cite{Zwolak2009} formulated a dehydration energy model by defining the energy of the hydration layer based on the Born equation and introducing a left fraction coefficient dependent on the nanotube's radius. As the nanotube radius diminishes, the left fraction coefficient decreases, indicating the degree of dehydration. Jiang et al.\cite{Shao2009a} proposed a hydration factor to account for the impact of ion-water orientation. An ion with a higher hydration factor corresponds to a higher shell order, which implies a smaller free energy difference. On the other hand, Fornasiero et al.\cite{Buchsbaum2021} introduced an effective enhancement factor to delineate the rapid permeation of certain ions in the nanotube. A larger effective enhancement factor correlates to a smaller free energy difference, resulting in accelerated ion permeation through the nanotube. However, it is vital to recognize that these models are fundamentally phenomenological and frequently necessitate the incorporation of empirical parameters. It should be noticed that continuum properties inside a nanotube, like relative permittivity\cite{Fumagalli2018,Zhang2023} and diffusion coefficient\cite{Li2023a}, deviate notably from their bulk counterparts. Determining these empirical parameters, not to mention the hydration structure in such confined spaces, presents a formidable challenge. While McClintock et al.\cite{Barabash2021} sought to introduce an atomic model to elucidate the hydration structure when an ion near a graphene surface using the Kirkwood superposition approximation\cite{Kirkwood1935b, Watanabe1960}, their model struggled to accurately predict the hydration structure in extreme narrow spaces. This limitation is inherent to the Kirkwood superposition approximation, which we will explore further in subsequent sections. Furthermore, the study did not establish a clear relationship between the hydration structure and the hydration free energy, further hindering its ability to describe its effect on ion transport properties. Given the array of limitations and challenges highlighted in the above studies, it's clear that current models fall short in accurately capturing the effect of the hydration structure change on ion transport. These shortcomings emphasize the pressing need for a more precise atomic model to truly capture these intricate dynamics.

\begin{figure}[htbp]
    \begin{center}
        \includegraphics[width=8cm]{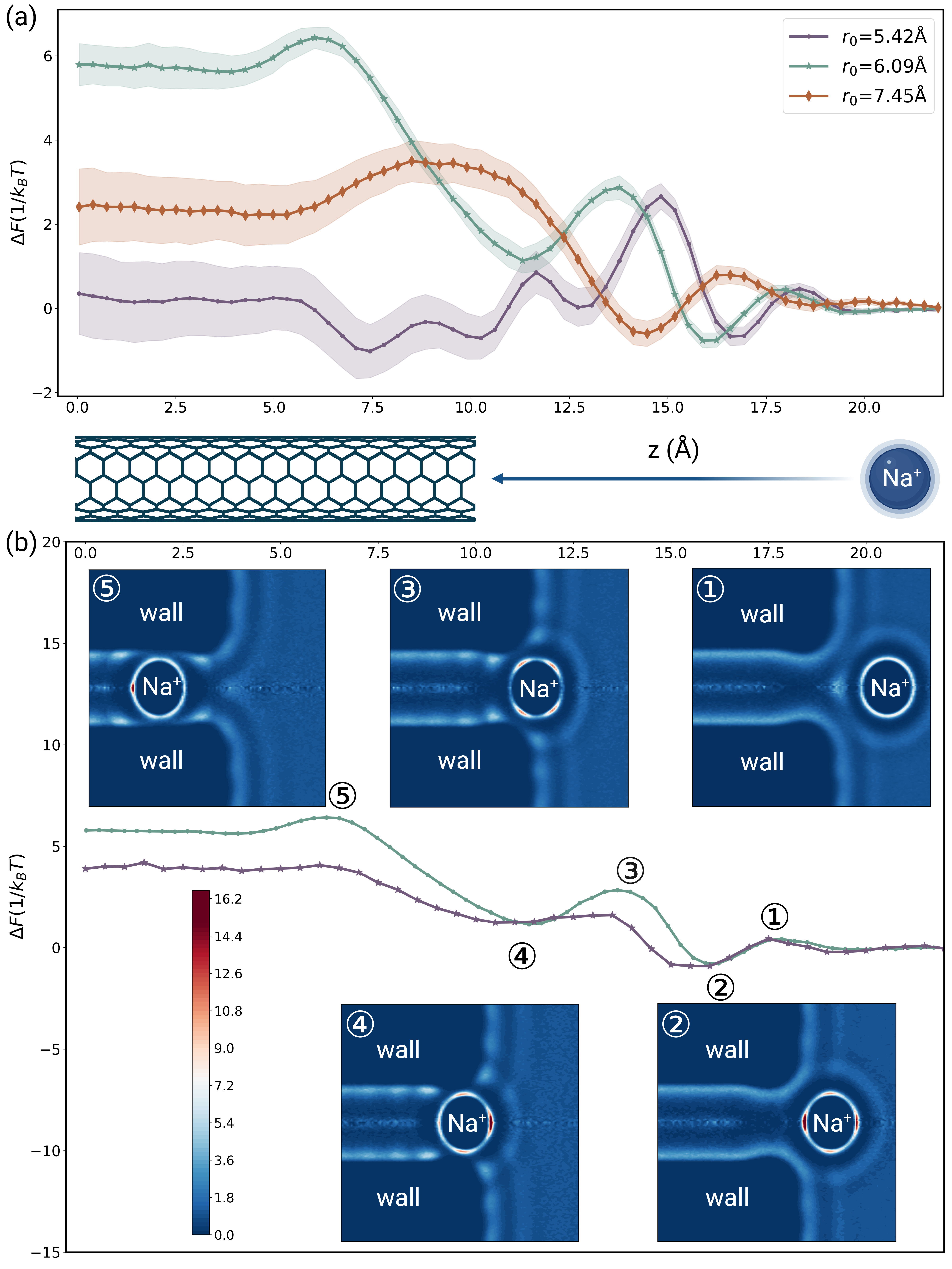}
    \end{center}
    \caption{A schematic illustration of the dehydration and hydration enhancement as an ion enters a carbon nanotube. (a) The free energy profile for a sodium ion entering carbon nanotubes of varying radii, with the bulk energy set to zero. Each profile's maximum, indicative of the hydration free energy barrier, does not monotonically increase with decreasing nanotube radius, which corroborates the complex dehydration principle. (b) provides five insets, labeled $\textcircled{1}$ to $\textcircled{5}$, showing the water distribution function derived from simulation at the peaks and valleys of the free energy profile when the radius equals $6.09\AA$. As the ion entering into the nanotube, its second hydration layer is completely stripped, while the first hydration layer alternately undergoes dehydration and hydration enhancement. This behavior generates the peaks and valleys observed in the free energy profile. The purple line, calculated using the water distribution obtained from the simulation, having the similar tendency with the reference free energy, represented by the teal line. This concurrence validates the accuracy of equation \eqref{equation-hydration-energy}.}
    \label{figure-02}
\end{figure}

% method

In this article, the umbrella sampling method is employed to extract the free energy profile of ion entry into carbon nanotubes through all-atom molecular dynamics (MD) simulations. We discovered that for nanotubes with radii less than $1\text{nm}$, the energy barrier, defined as the difference between the bulk energy and the maximum point on the free energy profile, does not consistently increase as the nanotube radius decreases (refer to Figure \ref{figure-02} (a)). Additionally, we identified multiple peaks and valleys in the free energy profile for all nanotubes considered in our study. These two findings challenge the prevailing view \cite{Song2009a,Richards2012}, which suggests that an ion simply experiences increased dehydration upon entering the carbon nanotube, leading to a monotonically increasing free energy profile. Therefore, in the case of nanotubes with small radius, there is an unknown mechanism by which ions are accommodated.
To delve deeper into this unknown  mechanism, we examined the water distribution function specifically when the ion locates at positions corresponding to each peak and valley, using data from our MD simulations, as illustrated in Figure \ref{figure-02} (b). Our analysis demonstrated that as an ion enter the nanotube, it experiences both dehydration and hydration enhancement within different hydration layers. Moreover, the valleys of the free energy profile align precisely with the hydration enhancement, while, in contrast, the peaks correspond precisely to the dehydration.
Inspired by these findings, we developed our model starting with the relationship between the water's translational free energy and the water distribution function\cite{Chandler1987}:

\begin{equation}
    F_{\text{trans}}^{(i)}(\mathbf{r}) = -k_BT\ln g^{(i)}(\mathbf{r})
    \label{equation-rdf-energy}
\end{equation}
In equation \eqref{equation-rdf-energy}, the terms $F^{(i)}_\text{trans}$ and $g^{(i)}$ represent the water's translational free energy and the water distribution function at position $\mathbf{r}$ with the presence of object $i$, respectively. 
Based on these two functions, the hydration energy of object $i$ is determined by aggregating the translational free energies of all water molecules, represented as: $F^{(i)}_{\text{hyd}} =\rho_0\int g^{(i)}(\mathbf{r})F_{\text{trans}}^{(i)}(\mathbf{r})\text{d}\mathbf{r}$. Here, the term $\rho_0$ is the number density of water molecules in the bulk condition.
Additionally, when a new object $j$ is introduced to the system, there is a redistribution of the water molecules, leading to a change in hydration free energy of object $i$. We define the difference in hydration free energy as:

\begin{equation}
    \Delta F_{\text{hyd}}^{(i)} =
    \rho_0\int \left[g^{(i, j)}(\mathbf{r}) - g^{(i)}(\mathbf{r})\right]
    F_{\text{trans}}^{(i)}(\mathbf{r})\text{d}\mathbf{r}
    \label{equation-hydration-energy}
\end{equation}
Here, $g^{(i, j)}\left(\textbf{r}\right)$ denotes the water distribution function in the presence of both objects $i$ and object $j$. To validate equation \eqref{equation-hydration-energy}, we use the water distribution function obtained from MD simulations to calculate the ion's hydration free energy profile during its entry into the carbon nanotube. The outcomes can be seen in Figure \ref{figure-02} (b), which aligns well with the actual free energy profile. This indicates that the free energy variation when an ion enters the carbon nanotube is primarily incurred by alterations in the hydration structure. Moreover, the free energy variation can be accurately characterized through a precise estimation of $g^{(i, j)}(\mathbf{r})$.

\begin{figure}[htbp]
    \begin{center}
        \includegraphics[width=8cm]{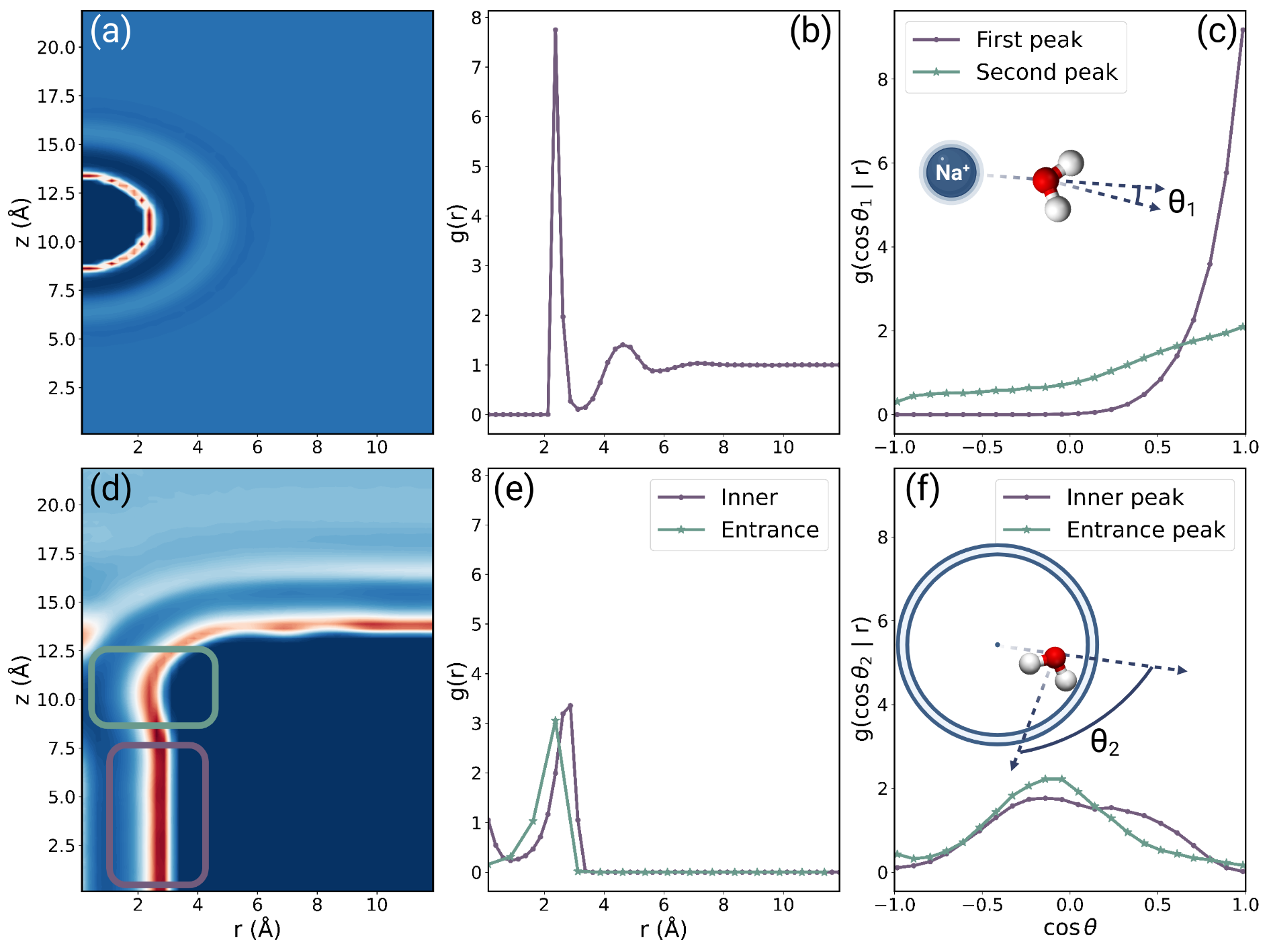}
    \end{center}
    \caption{
        A schematic illustration of the water distribution function and the water orientation distribution function of the sodium ion (first row) and the carbon nanotube with a radius of $6.09\AA$ (second row). Panels (a) and (d) depict the two-dimensional water distribution function, clearly illustrating the complex hydration structure near the nanotube's entry point, which gives rise to an intricate dehydration process in this region. Panels (b) and (e) represent the one-dimensional water distribution function, revealing the overlap of the first hydration layers of the ion and the carbon nanotube, both at the entrance and inside the nanotube. Panels (c) and (f) show the water orientation distribution function. The water molecules in the sodium ion's first hydration layer exhibit high orientation, with oxygen atoms pointing toward the ion, whereas this effect is substantially less pronounced in the second hydration layer. For the carbon nanotube, both at the entrance and inside, water molecules display different orientational preferences, with one hydrogen atom pointing toward the nanotube's wall, contrasting with the orientation around the sodium ion. These variations in orientational preferences contribute to the observed orientational competition.}
    \label{figure-03}
\end{figure}

In order to estimate $g^{(i, j)}(\mathbf{r})$, we refer to the Kirkwood superposition model \cite{Kirkwood1935b, Watanabe1960}. According to this model, we assume that the translational free energy of water, in the concurrent presence of both objects, is the superposition of the water's translational free energies when each object exists independently, devoid of the other's presence:

\begin{equation}
    g^{(i, j)}(\mathbf{r}) = g^{(i)}(\mathbf{r})g^{(j)}(\mathbf{r})
    \label{equation-superposition}
\end{equation}
However, when considering a sodium particle situated inside a nanotube with a radius of $6.09\AA$, equation \eqref{equation-superposition} suggests that $g^{(i, j)}(\mathbf{r})$ in overlapping regions surpasses a value of $28$ as the first peak value of the water distribution function for sodium and carbon nanotube is approximately $8$ and $3.5$, respectively (refer to Figure \ref{figure-03}, middel column).
Such a high value for the water distribution function is unrealistic, suggesting an overly dense congregation of water molecules. This phenomenon arises because the Kirkwood superposition model ignore the many-body effect stemming from the excess water molecules incurred by the presence of a new object. To rectify this, a cross-term representing the excess repulsive free energy should be integrated into the Kirkwood superposition model. As the water exhibits distinct orientational preferences when interacting with different objects (refer to Figure \ref{figure-03}, right column), there is an orientational competition during the redistribution of water molecules to accommodate a new object, resulting in an excess free energy. Therefore, we incorporating an orientational free energy term\cite{Lazaridis1996a} into the Kirkwood superposition model to account for the many-body effect:

\begin{equation}
    F_{\text{or}}^{(i)}(\theta | \mathbf{r}) = -k_BT\ln g^{(i)}(\theta | \mathbf{r})
    \label{equation-odf-energy}
\end{equation}
Here, $g^{(i)}(\theta|\mathbf{r})$ represents the conditional distribution defined by the joint distribution $g^{(i)}(\mathbf{r}, \theta)$ and the marginal distribution $g^{(i)}(\mathbf{r})$:
\begin{equation}
    g^{(i)}(\theta | \mathbf{r}) = \frac{g^{(i)}(\mathbf{r}, \theta)}{g^{(i)}(\mathbf{r})}\ , \
    g^{(i)}(\mathbf{r}) = \frac{1}{\Theta} \int_{\Theta} g^{(i)}(\mathbf{r}, \theta)
    \label{equation-odf}
\end{equation}
To estimate the joint distribution $g^{(i)}(\mathbf{r}, \theta)$, we extend the energy superposition assumption to the orientational space, similar to equation \eqref{equation-superposition}:
\begin{equation}
    g^{(i, j)}(\mathbf{r}, \theta)
    = g^{(i)}(\mathbf{r})g^{(j)}(\mathbf{r})
    g^{(i)}(\theta | \mathbf{r})g^{(j)}(\theta | \mathbf{r})
    \label{equation-grw}
\end{equation}
By considering the relationship between the joint distribution $g^{(i)}(\mathbf{r}, \theta)$ and the marginal distribution $g^{(i)}(\mathbf{r})$, we propose a modified superposition model for estimating the water distribution function:
\begin{equation}
    \begin{array}{rcl}
        \displaystyle g^{(i, j)}(\mathbf{r}) & = & \displaystyle \frac{1}{\Theta}\int_\Theta g^{(i)}(\mathbf{r})g^{(j)}(\mathbf{r})g^{(i)}(\theta | \mathbf{r})g^{(j)}(\theta | \mathbf{r}) \\\\
                                             & = & \displaystyle w^{(i, j)}(\mathbf{r})g^{(i)}(\mathbf{r})g^{(j)}(\mathbf{r})
    \end{array}
    \label{equation-modified-superposition}
\end{equation}
Where $w^{(i, j)}(\mathbf{r})$ is an orientational weight to account for the many-body effect from the viewpoint of orientational competition:

\begin{equation}
    w^{(i, j)}(\mathbf{r}) =
    \frac{1}{\Theta}\int_{\Theta}
    g^{(i)}(\theta | \mathbf{r})g^{(j)}(\theta | \mathbf{r})
    \text{d}\theta
    \label{equation-orientation-weight}
\end{equation}

\begin{figure}[htbp]
    \begin{center}
        \includegraphics[width=8cm]{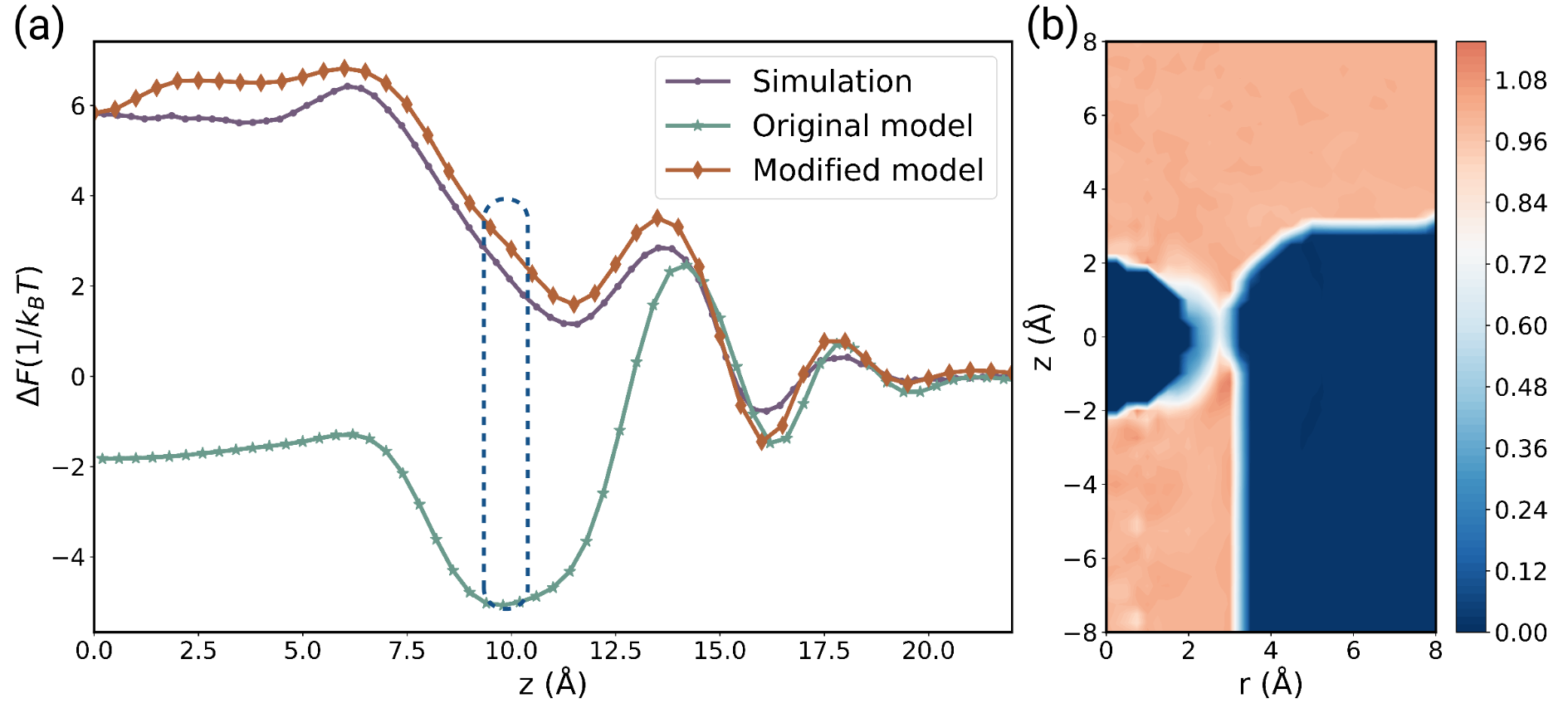}
    \end{center}
    \caption{This figure validates the predictive capability of the modified superposition models for hydration free energy profiles. Panel (a) demonstrates that by incorporating an orientational weight term into the original superposition model, the hydration free energy profile (orange line) closely aligns with the simulation results (purple line). This contrasts with the substantial deviation seen when using the original superposition model. Panel (b) displays the orientational weight term, highlighting the suppression effect in the overlap area and successfully representing the behavior of orientational competition.}
    \label{figure-04}
\end{figure}

% Result

In order to validate our modified superposition model, we contrast the hydration  free energy profiles calculated using equation \eqref{equation-hydration-energy} with $g^{(i, j)}(\mathbf{r})$ predicted by both the original and modified superposition models against simulation results. As demonstrated in Figure \ref{figure-04} (a), the modified superposition model shows robust alignment with the simulation results, whereas the original model displays substantial deviation beyond the second peak of the hydration free energy profile.
For a more comprehensive understanding of these divergences and agreements, we delve into the orientational weight $w^{(i, j)}(\mathbf{r})$ especially as ion located at the entrance area where the divergence in the hydration free energy profile is the most significant (refer to Figure \ref{figure-04} (b)). We identify considerable suppression in the overlap area. Consequently, we believe that the orientational weight aptly characterizes the orientational competition. Meanwhile, the orientational weight term converges rapidly to 1 in areas beyond the first hydration layer, affirming the short-range nature of water's orientational order within the hydration layer.

% Conclusion

\begin{equation}
    \left\{
    \begin{array}{l}
        \displaystyle g_{\text{original}}^{(i, j)}\left(\mathbf{r}\right) = 
        \frac{1}{\Theta^2}\int_{\Theta} g^{(i)}(r, \theta)\mathrm{d}\theta
        \int_{\Theta} g^{(j)}(r, \theta)\mathrm{d}\theta \\\\
        \displaystyle g_{\text{modified}}^{(i, j)}\left(\mathbf{r}\right) = \frac{1}{\Theta}\int_{\Theta} g^{(i)}(r, \theta) g^{(j)}(r, \theta)\mathrm{d}\theta
    \end{array}
    \right.
    \label{equation-model-comparision}
\end{equation}

To gain a deeper insight into the functioning of $w^{(i, j)}(\mathbf{r})$, we contrasted the original superposition model with the modified one, as outlined in equation \eqref{equation-model-comparision}.
The distinct difference between the original and the modified superposition models is evident in how they handle the integration in the orientational space.
The separated integration in the original model results in its failure to accurately capture orientational competition, as it decouples the influences from different objects on the same water molecule. 
Essentially, in each point of the spatial space, there are two distinct water molecules that independently interact with two objects and their hydration structures, instead of one single water molecule that melds the effects of both objects. 
In this regard, it's the ability of the orientational weight $w^{(i, j)}(\mathbf{r})$ to successfully couple the interactions from both objects and their hydration layers in the orientational space that allows it to aptly account for the many-body effect arising from the excess water molecules.

In conclusion, the findings of this study highlight the intricate process underlying the ion transport in confined environments, such as carbon nanotubes with radii less than 1 nm. This dynamic process involves both dehydration and hydration enhancement phenomena, which collectively contribute to the unique transport properties observed in such systems. Importantly, we found that the interaction of hydration layers of the ion and the nanotube gives rise to complex dynamics, leading to the formation of multiple peaks and valleys in the hydration free energy profile of ion entry into the nanotube.
Our work reveals the shortcomings of current models, which offer a simplified portrayal of these dynamics, often failing to fully encapsulate the complexity inherent in the dynamic process. By introducing an orientational weighting term to our modified superposition model, we've made a substantial leap forward in predicting accuracy, particularly in smaller nanotubes where the overlapping of the first hydration layers of the ion and the carbon nanotube occurs. This addition allows the model to account for the effects of orientational competition arising from the overlapping first hydration layers, a phenomenon largely neglected in prevailing models. The validation of our enhanced model through all-atom MD simulations corroborates its efficacy in capturing the intricate dynamics involving dehydration, hydration enhancement, and orientational aspects.

However, it is important to note that our study is just the beginning of a long scientific journey. In order to make our model more widely applicable, further refinement and generalization are necessary. One specific area that requires improvement is the estimation of the water orientation distribution function for different objects. This function plays a critical role in our model and obtaining an accurate estimate is crucial. Therefore, developing a highly efficient sampling method to obtain the water orientation distribution function will be essential for the further implementation of our model.
Moreover, in order to gain a comprehensive understanding of ion transport within confined systems, it is necessary to undertake more comprehensive studies. This should involve considering various environmental conditions, such as charged nanopores, which can significantly affect ion behavior. Additionally, incorporating more complex molecular structures into our model would provide a more realistic representation of the system under study.

By conducting such research efforts, we can deepen our understanding of the behavior of atoms and molecules in confined environments. This, in turn, will pave the way for the development of more precise and encompassing models in the future. Consequently, our research serves as a promising stride towards uncovering the intricacies of ion transport and lays the foundation for the evolution of more advanced models in this field.

\begin{acknowledgments}
    This is acknowledgments This is acknowledgments This is acknowledgments This is acknowledgments This is acknowledgments This is acknowledgments This is acknowledgments
\end{acknowledgments}

\bibliographystyle{apsrev4-1}
\bibliography{ref.bib} % The reference file name

\begin{thebibliography}{27}%
\makeatletter
\providecommand \@ifxundefined [1]{%
 \@ifx{#1\undefined}
}%
\providecommand \@ifnum [1]{%
 \ifnum #1\expandafter \@firstoftwo
 \else \expandafter \@secondoftwo
 \fi
}%
\providecommand \@ifx [1]{%
 \ifx #1\expandafter \@firstoftwo
 \else \expandafter \@secondoftwo
 \fi
}%
\providecommand \natexlab [1]{#1}%
\providecommand \enquote  [1]{``#1''}%
\providecommand \bibnamefont  [1]{#1}%
\providecommand \bibfnamefont [1]{#1}%
\providecommand \citenamefont [1]{#1}%
\providecommand \href@noop [0]{\@secondoftwo}%
\providecommand \href [0]{\begingroup \@sanitize@url \@href}%
\providecommand \@href[1]{\@@startlink{#1}\@@href}%
\providecommand \@@href[1]{\endgroup#1\@@endlink}%
\providecommand \@sanitize@url [0]{\catcode `\\12\catcode `\$12\catcode `\&12\catcode `\#12\catcode `\^12\catcode `\_12\catcode `\%12\relax}%
\providecommand \@@startlink[1]{}%
\providecommand \@@endlink[0]{}%
\providecommand \url  [0]{\begingroup\@sanitize@url \@url }%
\providecommand \@url [1]{\endgroup\@href {#1}{\urlprefix }}%
\providecommand \urlprefix  [0]{URL }%
\providecommand \Eprint [0]{\href }%
\providecommand \doibase [0]{http://dx.doi.org/}%
\providecommand \selectlanguage [0]{\@gobble}%
\providecommand \bibinfo  [0]{\@secondoftwo}%
\providecommand \bibfield  [0]{\@secondoftwo}%
\providecommand \translation [1]{[#1]}%
\providecommand \BibitemOpen [0]{}%
\providecommand \bibitemStop [0]{}%
\providecommand \bibitemNoStop [0]{.\EOS\space}%
\providecommand \EOS [0]{\spacefactor3000\relax}%
\providecommand \BibitemShut  [1]{\csname bibitem#1\endcsname}%
\let\auto@bib@innerbib\@empty
%</preamble>
\bibitem [{\citenamefont {Kavokine}\ \emph {et~al.}()\citenamefont {Kavokine}, \citenamefont {Netz},\ and\ \citenamefont {Bocquet}}]{Kavokine2021}%
  \BibitemOpen
  \bibfield  {author} {\bibinfo {author} {\bibfnamefont {N.}~\bibnamefont {Kavokine}}, \bibinfo {author} {\bibfnamefont {R.~R.}\ \bibnamefont {Netz}}, \ and\ \bibinfo {author} {\bibfnamefont {L.}~\bibnamefont {Bocquet}},\ }\href {\doibase 10.1146/annurev-fluid-071320-095958} {\ \textbf {\bibinfo {volume} {53}},\ \bibinfo {pages} {377}}\BibitemShut {NoStop}%
\bibitem [{\citenamefont {Chmiola}\ \emph {et~al.}()\citenamefont {Chmiola}, \citenamefont {Yushin}, \citenamefont {Gogotsi}, \citenamefont {Portet}, \citenamefont {Simon},\ and\ \citenamefont {Taberna}}]{Chmiola2006a}%
  \BibitemOpen
  \bibfield  {author} {\bibinfo {author} {\bibfnamefont {J.}~\bibnamefont {Chmiola}}, \bibinfo {author} {\bibfnamefont {G.}~\bibnamefont {Yushin}}, \bibinfo {author} {\bibfnamefont {Y.}~\bibnamefont {Gogotsi}}, \bibinfo {author} {\bibfnamefont {C.}~\bibnamefont {Portet}}, \bibinfo {author} {\bibfnamefont {P.}~\bibnamefont {Simon}}, \ and\ \bibinfo {author} {\bibfnamefont {P.~L.}\ \bibnamefont {Taberna}},\ }\href {\doibase 10.1126/science.1132195} {\ \textbf {\bibinfo {volume} {313}},\ \bibinfo {pages} {1760}}\BibitemShut {NoStop}%
\bibitem [{\citenamefont {Zhang}\ and\ \citenamefont {Zhao}()}]{Zhang2009a}%
  \BibitemOpen
  \bibfield  {author} {\bibinfo {author} {\bibfnamefont {L.~L.}\ \bibnamefont {Zhang}}\ and\ \bibinfo {author} {\bibfnamefont {X.~S.}\ \bibnamefont {Zhao}},\ }\href {\doibase 10.1039/b813846j} {\ \textbf {\bibinfo {volume} {38}},\ \bibinfo {pages} {2520}}\BibitemShut {NoStop}%
\bibitem [{\citenamefont {Yang}\ \emph {et~al.}()\citenamefont {Yang}, \citenamefont {Tian}, \citenamefont {Yin}, \citenamefont {Cui}, \citenamefont {Qian},\ and\ \citenamefont {Wei}}]{Yang2019}%
  \BibitemOpen
  \bibfield  {author} {\bibinfo {author} {\bibfnamefont {Z.}~\bibnamefont {Yang}}, \bibinfo {author} {\bibfnamefont {J.}~\bibnamefont {Tian}}, \bibinfo {author} {\bibfnamefont {Z.}~\bibnamefont {Yin}}, \bibinfo {author} {\bibfnamefont {C.}~\bibnamefont {Cui}}, \bibinfo {author} {\bibfnamefont {W.}~\bibnamefont {Qian}}, \ and\ \bibinfo {author} {\bibfnamefont {F.}~\bibnamefont {Wei}},\ }\href {\doibase 10.1016/j.carbon.2018.10.010} {\ \textbf {\bibinfo {volume} {141}},\ \bibinfo {pages} {467}}\BibitemShut {NoStop}%
\bibitem [{\citenamefont {Wang}\ \emph {et~al.}()\citenamefont {Wang}, \citenamefont {Hou}, \citenamefont {Yu},\ and\ \citenamefont {Hou}}]{Wang2020}%
  \BibitemOpen
  \bibfield  {author} {\bibinfo {author} {\bibfnamefont {M.}~\bibnamefont {Wang}}, \bibinfo {author} {\bibfnamefont {Y.}~\bibnamefont {Hou}}, \bibinfo {author} {\bibfnamefont {L.}~\bibnamefont {Yu}}, \ and\ \bibinfo {author} {\bibfnamefont {X.}~\bibnamefont {Hou}},\ }\href {\doibase 10.1021/acs.nanolett.0c02999} {\ \textbf {\bibinfo {volume} {20}},\ \bibinfo {pages} {6937}}\BibitemShut {NoStop}%
\bibitem [{\citenamefont {{T. Mouterde}}\ \emph {et~al.}()\citenamefont {{T. Mouterde}}, \citenamefont {{Timothée Mouterde}}, \citenamefont {Mouterde}, \citenamefont {Keerthi}, \citenamefont {Poggioli}, \citenamefont {{Anthony R. Poggioli}}, \citenamefont {Dar}, \citenamefont {Siria}, \citenamefont {Geim}, \citenamefont {Bocquet},\ and\ \citenamefont {Radha}}]{T.Mouterde2019}%
  \BibitemOpen
  \bibfield  {author} {\bibinfo {author} {\bibnamefont {{T. Mouterde}}}, \bibinfo {author} {\bibnamefont {{Timothée Mouterde}}}, \bibinfo {author} {\bibfnamefont {T.}~\bibnamefont {Mouterde}}, \bibinfo {author} {\bibfnamefont {A.}~\bibnamefont {Keerthi}}, \bibinfo {author} {\bibfnamefont {A.~R.}\ \bibnamefont {Poggioli}}, \bibinfo {author} {\bibnamefont {{Anthony R. Poggioli}}}, \bibinfo {author} {\bibfnamefont {S.~A.}\ \bibnamefont {Dar}}, \bibinfo {author} {\bibfnamefont {A.}~\bibnamefont {Siria}}, \bibinfo {author} {\bibfnamefont {A.~K.}\ \bibnamefont {Geim}}, \bibinfo {author} {\bibfnamefont {L.}~\bibnamefont {Bocquet}}, \ and\ \bibinfo {author} {\bibfnamefont {B.}~\bibnamefont {Radha}},\ }\href {\doibase 10.1038/s41586-019-0961-5} {\ \textbf {\bibinfo {volume} {567}},\ \bibinfo {pages} {87}}\BibitemShut {NoStop}%
\bibitem [{\citenamefont {Lucas}\ \emph {et~al.}()\citenamefont {Lucas}, \citenamefont {Lin}, \citenamefont {Baker},\ and\ \citenamefont {Siwy}}]{Lucas2020}%
  \BibitemOpen
  \bibfield  {author} {\bibinfo {author} {\bibfnamefont {R.~A.}\ \bibnamefont {Lucas}}, \bibinfo {author} {\bibfnamefont {C.-Y.}\ \bibnamefont {Lin}}, \bibinfo {author} {\bibfnamefont {L.~A.}\ \bibnamefont {Baker}}, \ and\ \bibinfo {author} {\bibfnamefont {Z.~S.}\ \bibnamefont {Siwy}},\ }\href {\doibase 10.1038/s41467-020-15398-3} {\ \textbf {\bibinfo {volume} {11}},\ \bibinfo {pages} {1568}}\BibitemShut {NoStop}%
\bibitem [{\citenamefont {Xue}\ \emph {et~al.}()\citenamefont {Xue}, \citenamefont {Xia}, \citenamefont {Yang}, \citenamefont {Alsaid}, \citenamefont {Fong}, \citenamefont {Wang},\ and\ \citenamefont {Zhang}}]{Xue2021}%
  \BibitemOpen
  \bibfield  {author} {\bibinfo {author} {\bibfnamefont {Y.}~\bibnamefont {Xue}}, \bibinfo {author} {\bibfnamefont {Y.}~\bibnamefont {Xia}}, \bibinfo {author} {\bibfnamefont {S.}~\bibnamefont {Yang}}, \bibinfo {author} {\bibfnamefont {Y.}~\bibnamefont {Alsaid}}, \bibinfo {author} {\bibfnamefont {K.~Y.}\ \bibnamefont {Fong}}, \bibinfo {author} {\bibfnamefont {Y.}~\bibnamefont {Wang}}, \ and\ \bibinfo {author} {\bibfnamefont {X.}~\bibnamefont {Zhang}},\ }\href {\doibase 10.1126/science.abb5144} {\ \textbf {\bibinfo {volume} {372}},\ \bibinfo {pages} {501}}\BibitemShut {NoStop}%
\bibitem [{\citenamefont {Zwanzig}()}]{Zwanzig1963}%
  \BibitemOpen
  \bibfield  {author} {\bibinfo {author} {\bibfnamefont {R.}~\bibnamefont {Zwanzig}},\ }\href {\doibase 10.1063/1.1776929} {\ \textbf {\bibinfo {volume} {38}},\ \bibinfo {pages} {1603}}\BibitemShut {NoStop}%
\bibitem [{\citenamefont {Banerjee}\ and\ \citenamefont {Bagchi}()}]{Banerjee2019}%
  \BibitemOpen
  \bibfield  {author} {\bibinfo {author} {\bibfnamefont {P.}~\bibnamefont {Banerjee}}\ and\ \bibinfo {author} {\bibfnamefont {B.}~\bibnamefont {Bagchi}},\ }\href {\doibase 10.1063/1.5090765} {\ \textbf {\bibinfo {volume} {150}},\ \bibinfo {pages} {190901}}\BibitemShut {NoStop}%
\bibitem [{\citenamefont {Koga}\ \emph {et~al.}()\citenamefont {Koga}, \citenamefont {Gao}, \citenamefont {Tanaka},\ and\ \citenamefont {Zeng}}]{Koga2001}%
  \BibitemOpen
  \bibfield  {author} {\bibinfo {author} {\bibfnamefont {K.}~\bibnamefont {Koga}}, \bibinfo {author} {\bibfnamefont {G.~T.}\ \bibnamefont {Gao}}, \bibinfo {author} {\bibfnamefont {H.}~\bibnamefont {Tanaka}}, \ and\ \bibinfo {author} {\bibfnamefont {X.~C.}\ \bibnamefont {Zeng}},\ }\href {\doibase 10.1038/35090532} {\ \textbf {\bibinfo {volume} {412}},\ \bibinfo {pages} {802}}\BibitemShut {NoStop}%
\bibitem [{\citenamefont {He}\ \emph {et~al.}()\citenamefont {He}, \citenamefont {Zhou}, \citenamefont {Lu},\ and\ \citenamefont {Corry}}]{He2013a}%
  \BibitemOpen
  \bibfield  {author} {\bibinfo {author} {\bibfnamefont {Z.}~\bibnamefont {He}}, \bibinfo {author} {\bibfnamefont {J.}~\bibnamefont {Zhou}}, \bibinfo {author} {\bibfnamefont {X.}~\bibnamefont {Lu}}, \ and\ \bibinfo {author} {\bibfnamefont {B.}~\bibnamefont {Corry}},\ }\href {\doibase 10.1021/jp4025206} {\ \textbf {\bibinfo {volume} {117}},\ \bibinfo {pages} {11412}}\BibitemShut {NoStop}%
\bibitem [{\citenamefont {Dalla~Bernardina}\ \emph {et~al.}()\citenamefont {Dalla~Bernardina}, \citenamefont {Paineau}, \citenamefont {Brubach}, \citenamefont {Judeinstein}, \citenamefont {Rouzière}, \citenamefont {Launois},\ and\ \citenamefont {Roy}}]{DallaBernardina2016}%
  \BibitemOpen
  \bibfield  {author} {\bibinfo {author} {\bibfnamefont {S.}~\bibnamefont {Dalla~Bernardina}}, \bibinfo {author} {\bibfnamefont {E.}~\bibnamefont {Paineau}}, \bibinfo {author} {\bibfnamefont {J.-B.}\ \bibnamefont {Brubach}}, \bibinfo {author} {\bibfnamefont {P.}~\bibnamefont {Judeinstein}}, \bibinfo {author} {\bibfnamefont {S.}~\bibnamefont {Rouzière}}, \bibinfo {author} {\bibfnamefont {P.}~\bibnamefont {Launois}}, \ and\ \bibinfo {author} {\bibfnamefont {P.}~\bibnamefont {Roy}},\ }\href {\doibase 10.1021/jacs.6b02635} {\ \textbf {\bibinfo {volume} {138}},\ \bibinfo {pages} {10437}}\BibitemShut {NoStop}%
\bibitem [{\citenamefont {Shao}\ \emph {et~al.}()\citenamefont {Shao}, \citenamefont {Zhou}, \citenamefont {Lu}, \citenamefont {Lu}, \citenamefont {Zhu},\ and\ \citenamefont {Jiang}}]{Shao2009a}%
  \BibitemOpen
  \bibfield  {author} {\bibinfo {author} {\bibfnamefont {Q.}~\bibnamefont {Shao}}, \bibinfo {author} {\bibfnamefont {J.}~\bibnamefont {Zhou}}, \bibinfo {author} {\bibfnamefont {L.}~\bibnamefont {Lu}}, \bibinfo {author} {\bibfnamefont {X.}~\bibnamefont {Lu}}, \bibinfo {author} {\bibfnamefont {Y.}~\bibnamefont {Zhu}}, \ and\ \bibinfo {author} {\bibfnamefont {S.}~\bibnamefont {Jiang}},\ }\href {\doibase 10.1021/nl803044k} {\ \textbf {\bibinfo {volume} {9}},\ \bibinfo {pages} {989}}\BibitemShut {NoStop}%
\bibitem [{\citenamefont {Zwolak}\ \emph {et~al.}()\citenamefont {Zwolak}, \citenamefont {Lagerqvist},\ and\ \citenamefont {Di~Ventra}}]{Zwolak2009}%
  \BibitemOpen
  \bibfield  {author} {\bibinfo {author} {\bibfnamefont {M.}~\bibnamefont {Zwolak}}, \bibinfo {author} {\bibfnamefont {J.}~\bibnamefont {Lagerqvist}}, \ and\ \bibinfo {author} {\bibfnamefont {M.}~\bibnamefont {Di~Ventra}},\ }\href {\doibase 10.1103/PhysRevLett.103.128102} {\ \textbf {\bibinfo {volume} {103}},\ \bibinfo {pages} {128102}}\BibitemShut {NoStop}%
\bibitem [{\citenamefont {Sahu}\ \emph {et~al.}()\citenamefont {Sahu}, \citenamefont {Di~Ventra},\ and\ \citenamefont {Zwolak}}]{Sahu2017c}%
  \BibitemOpen
  \bibfield  {author} {\bibinfo {author} {\bibfnamefont {S.}~\bibnamefont {Sahu}}, \bibinfo {author} {\bibfnamefont {M.}~\bibnamefont {Di~Ventra}}, \ and\ \bibinfo {author} {\bibfnamefont {M.}~\bibnamefont {Zwolak}},\ }\href {\doibase 10.1021/acs.nanolett.7b01399} {\ \textbf {\bibinfo {volume} {17}},\ \bibinfo {pages} {4719}}\BibitemShut {NoStop}%
\bibitem [{\citenamefont {Buchsbaum}\ \emph {et~al.}()\citenamefont {Buchsbaum}, \citenamefont {Jue}, \citenamefont {Sawvel}, \citenamefont {Chen}, \citenamefont {Meshot}, \citenamefont {Park}, \citenamefont {Wood}, \citenamefont {Wu}, \citenamefont {Bilodeau}, \citenamefont {Aydin}, \citenamefont {Pham}, \citenamefont {Lau},\ and\ \citenamefont {Fornasiero}}]{Buchsbaum2021}%
  \BibitemOpen
  \bibfield  {author} {\bibinfo {author} {\bibfnamefont {S.~F.}\ \bibnamefont {Buchsbaum}}, \bibinfo {author} {\bibfnamefont {M.~L.}\ \bibnamefont {Jue}}, \bibinfo {author} {\bibfnamefont {A.~M.}\ \bibnamefont {Sawvel}}, \bibinfo {author} {\bibfnamefont {C.}~\bibnamefont {Chen}}, \bibinfo {author} {\bibfnamefont {E.~R.}\ \bibnamefont {Meshot}}, \bibinfo {author} {\bibfnamefont {S.~J.}\ \bibnamefont {Park}}, \bibinfo {author} {\bibfnamefont {M.}~\bibnamefont {Wood}}, \bibinfo {author} {\bibfnamefont {K.~J.}\ \bibnamefont {Wu}}, \bibinfo {author} {\bibfnamefont {C.~L.}\ \bibnamefont {Bilodeau}}, \bibinfo {author} {\bibfnamefont {F.}~\bibnamefont {Aydin}}, \bibinfo {author} {\bibfnamefont {T.~A.}\ \bibnamefont {Pham}}, \bibinfo {author} {\bibfnamefont {E.~Y.}\ \bibnamefont {Lau}}, \ and\ \bibinfo {author} {\bibfnamefont {F.}~\bibnamefont {Fornasiero}},\ }\href {\doibase 10.1002/advs.202001802} {\ \textbf {\bibinfo {volume} {8}},\ \bibinfo {pages} {2001802}}\BibitemShut {NoStop}%
\bibitem [{\citenamefont {Fumagalli}\ \emph {et~al.}()\citenamefont {Fumagalli}, \citenamefont {Esfandiar}, \citenamefont {Fabregas}, \citenamefont {Hu}, \citenamefont {Ares}, \citenamefont {Janardanan}, \citenamefont {Yang}, \citenamefont {Radha}, \citenamefont {Taniguchi}, \citenamefont {Watanabe}, \citenamefont {Gomila}, \citenamefont {Novoselov},\ and\ \citenamefont {Geim}}]{Fumagalli2018}%
  \BibitemOpen
  \bibfield  {author} {\bibinfo {author} {\bibfnamefont {L.}~\bibnamefont {Fumagalli}}, \bibinfo {author} {\bibfnamefont {A.}~\bibnamefont {Esfandiar}}, \bibinfo {author} {\bibfnamefont {R.}~\bibnamefont {Fabregas}}, \bibinfo {author} {\bibfnamefont {S.}~\bibnamefont {Hu}}, \bibinfo {author} {\bibfnamefont {P.}~\bibnamefont {Ares}}, \bibinfo {author} {\bibfnamefont {A.}~\bibnamefont {Janardanan}}, \bibinfo {author} {\bibfnamefont {Q.}~\bibnamefont {Yang}}, \bibinfo {author} {\bibfnamefont {B.}~\bibnamefont {Radha}}, \bibinfo {author} {\bibfnamefont {T.}~\bibnamefont {Taniguchi}}, \bibinfo {author} {\bibfnamefont {K.}~\bibnamefont {Watanabe}}, \bibinfo {author} {\bibfnamefont {G.}~\bibnamefont {Gomila}}, \bibinfo {author} {\bibfnamefont {K.~S.}\ \bibnamefont {Novoselov}}, \ and\ \bibinfo {author} {\bibfnamefont {A.~K.}\ \bibnamefont {Geim}},\ }\href {\doibase 10.1126/science.aat4191} {\ \textbf {\bibinfo {volume} {360}},\ \bibinfo {pages} {1339}}\BibitemShut {NoStop}%
\bibitem [{\citenamefont {Zhang}\ \emph {et~al.}()\citenamefont {Zhang}, \citenamefont {Yue}, \citenamefont {Panagiotopoulos}, \citenamefont {Klein},\ and\ \citenamefont {Wu}}]{Zhang2023}%
  \BibitemOpen
  \bibfield  {author} {\bibinfo {author} {\bibfnamefont {C.}~\bibnamefont {Zhang}}, \bibinfo {author} {\bibfnamefont {S.}~\bibnamefont {Yue}}, \bibinfo {author} {\bibfnamefont {A.~Z.}\ \bibnamefont {Panagiotopoulos}}, \bibinfo {author} {\bibfnamefont {M.~L.}\ \bibnamefont {Klein}}, \ and\ \bibinfo {author} {\bibfnamefont {X.}~\bibnamefont {Wu}},\ }\href {\doibase 10.1103/PhysRevLett.131.076801} {\ \textbf {\bibinfo {volume} {131}},\ \bibinfo {pages} {076801}}\BibitemShut {NoStop}%
\bibitem [{\citenamefont {Li}\ \emph {et~al.}()\citenamefont {Li}, \citenamefont {Misra}, \citenamefont {Li}, \citenamefont {Yao}, \citenamefont {Zhao}, \citenamefont {Zhang}, \citenamefont {Chen}, \citenamefont {Blankschtein},\ and\ \citenamefont {Noy}}]{Li2023a}%
  \BibitemOpen
  \bibfield  {author} {\bibinfo {author} {\bibfnamefont {Z.}~\bibnamefont {Li}}, \bibinfo {author} {\bibfnamefont {R.~P.}\ \bibnamefont {Misra}}, \bibinfo {author} {\bibfnamefont {Y.}~\bibnamefont {Li}}, \bibinfo {author} {\bibfnamefont {Y.-C.}\ \bibnamefont {Yao}}, \bibinfo {author} {\bibfnamefont {S.}~\bibnamefont {Zhao}}, \bibinfo {author} {\bibfnamefont {Y.}~\bibnamefont {Zhang}}, \bibinfo {author} {\bibfnamefont {Y.}~\bibnamefont {Chen}}, \bibinfo {author} {\bibfnamefont {D.}~\bibnamefont {Blankschtein}}, \ and\ \bibinfo {author} {\bibfnamefont {A.}~\bibnamefont {Noy}},\ }\href {\doibase 10.1038/s41565-022-01276-0} {\ \textbf {\bibinfo {volume} {18}},\ \bibinfo {pages} {177}}\BibitemShut {NoStop}%
\bibitem [{\citenamefont {Barabash}\ \emph {et~al.}()\citenamefont {Barabash}, \citenamefont {Gibby}, \citenamefont {Guardiani}, \citenamefont {Smolyanitsky}, \citenamefont {Luchinsky},\ and\ \citenamefont {McClintock}}]{Barabash2021}%
  \BibitemOpen
  \bibfield  {author} {\bibinfo {author} {\bibfnamefont {M.~L.}\ \bibnamefont {Barabash}}, \bibinfo {author} {\bibfnamefont {W.~A.~T.}\ \bibnamefont {Gibby}}, \bibinfo {author} {\bibfnamefont {C.}~\bibnamefont {Guardiani}}, \bibinfo {author} {\bibfnamefont {A.}~\bibnamefont {Smolyanitsky}}, \bibinfo {author} {\bibfnamefont {D.~G.}\ \bibnamefont {Luchinsky}}, \ and\ \bibinfo {author} {\bibfnamefont {P.~V.~E.}\ \bibnamefont {McClintock}},\ }\href {\doibase 10.1038/s43246-021-00162-x} {\ \textbf {\bibinfo {volume} {2}},\ \bibinfo {pages} {65}}\BibitemShut {NoStop}%
\bibitem [{\citenamefont {Kirkwood}()}]{Kirkwood1935b}%
  \BibitemOpen
  \bibfield  {author} {\bibinfo {author} {\bibfnamefont {J.~G.}\ \bibnamefont {Kirkwood}},\ }\href {\doibase 10.1063/1.1749657} {\ \textbf {\bibinfo {volume} {3}},\ \bibinfo {pages} {300}}\BibitemShut {NoStop}%
\bibitem [{\citenamefont {Watanabe}()}]{Watanabe1960}%
  \BibitemOpen
  \bibfield  {author} {\bibinfo {author} {\bibfnamefont {S.}~\bibnamefont {Watanabe}},\ }\href {\doibase 10.1147/rd.41.0066} {\ \textbf {\bibinfo {volume} {4}},\ \bibinfo {pages} {66}}\BibitemShut {NoStop}%
\bibitem [{\citenamefont {Song}\ and\ \citenamefont {Corry}()}]{Song2009a}%
  \BibitemOpen
  \bibfield  {author} {\bibinfo {author} {\bibfnamefont {C.}~\bibnamefont {Song}}\ and\ \bibinfo {author} {\bibfnamefont {B.}~\bibnamefont {Corry}},\ }\href {\doibase 10.1021/jp810102u} {\ \textbf {\bibinfo {volume} {113}},\ \bibinfo {pages} {7642}}\BibitemShut {NoStop}%
\bibitem [{\citenamefont {Richards}\ \emph {et~al.}()\citenamefont {Richards}, \citenamefont {Schäfer}, \citenamefont {Richards},\ and\ \citenamefont {Corry}}]{Richards2012}%
  \BibitemOpen
  \bibfield  {author} {\bibinfo {author} {\bibfnamefont {L.~A.}\ \bibnamefont {Richards}}, \bibinfo {author} {\bibfnamefont {A.~I.}\ \bibnamefont {Schäfer}}, \bibinfo {author} {\bibfnamefont {B.~S.}\ \bibnamefont {Richards}}, \ and\ \bibinfo {author} {\bibfnamefont {B.}~\bibnamefont {Corry}},\ }\href {\doibase 10.1002/smll.201102056} {\ \textbf {\bibinfo {volume} {8}},\ \bibinfo {pages} {1701}}\BibitemShut {NoStop}%
\bibitem [{\citenamefont {Chandler}()}]{Chandler1987}%
  \BibitemOpen
  \bibfield  {author} {\bibinfo {author} {\bibfnamefont {D.}~\bibnamefont {Chandler}},\ }\href@noop {} {\emph {\bibinfo {title} {Introduction to Modern Statistical Mechanics}}}\ (\bibinfo  {publisher} {Oxford University Press})\BibitemShut {NoStop}%
\bibitem [{\citenamefont {Lazaridis}\ and\ \citenamefont {Karplus}()}]{Lazaridis1996a}%
  \BibitemOpen
  \bibfield  {author} {\bibinfo {author} {\bibfnamefont {T.}~\bibnamefont {Lazaridis}}\ and\ \bibinfo {author} {\bibfnamefont {M.}~\bibnamefont {Karplus}},\ }\href {\doibase 10.1063/1.472247} {\ \textbf {\bibinfo {volume} {105}},\ \bibinfo {pages} {4294}}\BibitemShut {NoStop}%
\end{thebibliography}%
\end{document}